\documentclass[pdflatex,sn-nature]{sn-jnl}

\usepackage{graphicx}%
\usepackage{multirow}%
\usepackage{amsmath,amssymb,amsfonts}%
\usepackage{amsthm}%
\usepackage{mathrsfs}%
\usepackage[title]{appendix}%
\usepackage{xcolor}%
\usepackage{textcomp}%
\usepackage{manyfoot}%
\usepackage{booktabs}%
\usepackage{algorithm}%
\usepackage{algorithmicx}%
\usepackage{algpseudocode}%
\usepackage{listings}%
\usepackage{aasmacros}
\usepackage{natbib}
\usepackage{mhchem}

\newcommand\subs[1]{\textsubscript{#1}}

\newcommand\rh[1]{\textcolor{black}{{\textit{r}\subs{H}}#1}}

\newcommand\trot[1]{\textcolor{black}{{\textit{T}\subs{rot}}#1}}

\newcommand\kms[1]{\textcolor{black}{{km\,s$^{-1}$}#1}}

\newcommand\ps[1]{\textcolor{black}{{s$^{-1}$}#1}}

\newcommand{\gtsim}{$\buildrel > \over \sim $}

\theoremstyle{thmstyleone}%
\theoremstyle{thmstyletwo}%

\theoremstyle{thmstylethree}%

\begin{document}

\title[Article Title]{An Enriched Methane D/H Ratio in the Interstellar Object 3I/ATLAS}


\author*[1,2]{\fnm{Nathan X.} \sur{Roth}}\email{nathaniel.x.roth@nasa.gov}

\author[1,3]{\fnm{Martin} \sur{Cordiner}}

\author[1]{\fnm{Stefanie} \sur{Milam}}

\author[1]{\fnm{Geronimo} \sur{Villanueva}}

\author[1]{\fnm{Steven} \sur{Charnley}}

\author[4]{Nicolas Biver}

\author[4]{Dominique Bockel{\'e}e-Morvan}

\author[5]{\fnm{Dennis} \sur{Bodewits}}

\author[4]{Jacques Crovisier}

\author[6,7]{\fnm{Maria N.} \sur{Drozdovskaya}}

\author[8]{Davide Farnocchia}

\author[9]{\fnm{Kenji} \sur{Furuya}}

\author[10]{\fnm{Michael S. P.} \sur{Kelley}}

\author[11]{\fnm{Marco} \sur{Micheli}}

\author[5]{\fnm{John W.} \sur{Noonan}}

\author[12]{Cyrielle Opitom}

\author[13]{Megan E. Schwamb}

\author[14]{Cristina A. Thomas}

\affil*[1]{\orgdiv{Solar System Exploration Division}, \orgname{NASA Goddard Space Flight Center}, \orgaddress{\street{8800 Greenbelt Rd}, \city{Greenbelt}, \postcode{20771}, \state{MD}, \country{USA}}}

\affil[2]{\orgdiv{Department of Physics}, \orgname{American University}, \orgaddress{\street{4400 Massachusetts Ave NW}, \city{Washington}, \postcode{20016}, \state{DC}, \country{USA}}}

\affil[3]{\orgdiv{Department of Physics}, \orgname{Catholic University of America}, \orgaddress{\street{620 Michigan Ave. NE}, \city{Washington}, \postcode{DC 20064}, \country{USA}}}

\affil[4]{LIRA, Observatoire de Paris, Universit\'e PSL, CNRS, Sorbonne Universit\'e, Universit\'e Paris Cit\'e, CY Cergy Paris Universit\'e, 5 place Jules Janssen, F-92190 Meudon, France}

\affil[5]{\orgdiv{Department of Physics}, \orgname{Auburn University}, \orgaddress{\street{Edmund C. Leach Science Center}, \postcode{36382},  \city{Auburn}, \state{AL}, \country{USA}}}

\affil[6]{\orgname{Physikalisch-Meteorologisches Observatorium Davos und Weltstrahlungszentrum (PMOD/WRC)}, \orgaddress{Dorfstrasse 33}, \postcode{7260} \city{Davos Dorf}, \country{Switzerland}}

\affil[7]{\orgdiv{Department of Chemistry, Biochemistry and Pharmaceutical Sciences (DCBP)}, \orgname{Universit{\"a}t Bern}, \orgaddress{Freiestrasse 3}, \postcode{3012}, \city{Bern}, \country{Switzerland}}

\affil[8]{Jet Propulsion Laboratory, California Institute of Technology, 4800 Oak Grove Dr., Pasadena, CA 91109, USA}

\affil[9]{\orgdiv{Pioneering Research Institute}, \orgname{RIKEN}, \orgaddress{2-1 Hirosawa}, \city{Wako-shi, Saitama}, \postcode{351-0198}, \country{Japan}}

\affil[10]{\orgdiv{Department of Astronomy}, \orgname{University of Maryland}, \orgaddress{\street{4296 Stadium Dr., PSC Bldg. 415}, \postcode{20742-2421},  \city{College Park}, \state{MD}, \country{USA}}}

\affil[11]{\orgdiv{ESA NEO Coordination Centre}, \orgname{Planetary Defence Office, European Space Agency}, \orgaddress{\street{Largo Galileo Galilei, 1}, \postcode{00044},  \city{Frascati}, \state{RM}, \country{Italy}}}

\affil[12]{Institute for Astronomy, University of Edinburgh, Royal Observatory, Edinburgh EH9 3HJ, UK}

\affil[13]{Astrophysics Research Centre, School of Mathematics and Physics, Queen's University Belfast, Belfast BT7 1NN, UK}

\affil[14]{Northern Arizona University, Department of Astronomy and Planetary Science, P.O. Box 6010, Flagstaff, AZ, 86011 USA}



\abstract{\unboldmath Interstellar objects are interlopers from other planetary systems, and their volatile compositions provide a glimpse into planet formation around their host star. We present near-infrared spectra of the coma of interstellar object 3I/ATLAS measured with the James Webb Space Telescope. Our results demonstrate an unexpectedly high $\mathrm{D}/\mathrm{H} = (3.33\pm0.31)\%$ for methane and represent an exceedingly rare detection of deuterated organic molecules in an interstellar object. This D/H ratio exceeds any other value for methane measured in the solar system, and is a factor of $14\pm2$ higher than that measured in comet 67P/Churyumov-Gerasimenko by the Rosetta spacecraft, the only other comet for which \ce{CH3D} has been detected. Both 3I/ATLAS and 67P/Churyumov-Gerasimenko show a higher degree of methane dueteration compared to water, consistent with trends seen in other solar system bodies and the nearby interstellar medium, where deuteration in organics exceeds that of water by up to an order of magnitude. The D/H ratio in methane itself is observationally unconstrained in extrasolar sources to date, but the enriched ratio in 3I/ATLAS is similar to those measured in methanol and formaldehyde toward primitive environments. The elevated D/H ratio for methane in 3I/ATLAS is consistent with the hypothesis that its ices formed in a cold, low metallicity interstellar or protostellar environment under a high cosmic ray irradiation rate \citep{Cordiner2026}.}




\maketitle
\begin{center}
\textit{\small Note: This is a pre-print. The version of record is published in Roth et al. 2026, \textit{Nature Astronomy}, doi: https://doi.org/10.1038/s41550-026-03002-5}
\end{center}

\section{Introduction}\label{sec1}

Reconstructing the physics and chemistry of planet formation is a fundamental pursuit in astronomy. Dynamical simulations of accretion and orbital architecture, returned samples of comets and asteroids, analysis of meteorites, and remote sensing studies of the present day solar system have all helped to inform our understanding of planet formation around the Sun \citep[e.g.,][ and references therein]{Gourier2008,Duprat2010,Levison2011,Ito2022,Miguel2023,Biver2024b}. Yet revealing the interiors of protoplanetary disks around other young stars is challenging. Owing to significant opacity, the midplane of the protoplanetary disk in extrasolar systems and the ice-phase chemistry occurring therein is generally not accessible via remote sensing. Thus, astrochemical models inform the majority of our knowledge of these environments in other stellar systems. Given their provenance from our own protoplanetary disk midplane, the compositions of solar system comets serve as one of the few observational constraints on these models \citep[e.g.,][]{Willacy2022}. 

The recent discovery of comet-like objects passing through the solar system on hyperbolic, gravitationally unbound trajectories, namely 2I/Borisov and 3I/ATLAS, has opened up a new window into the chemistry of planet formation around other stars by delivering their contents to our own planetary system for analysis \citep{Bailer-Jones2020,Seligman2025}. Dynamical and compositional studies of these enigmatic objects may provide clues to their origins. Soon after its discovery, multiple studies leveraged models of Galactic dynamics and chemistry to suggest 3I/ATLAS formed in a low-metallicity star system $>3$ Gyr ago given its radiant direction and high heliocentric velocity, although discerning a thick- vs. thin-disk origin was difficult \citep{Guo2025,Taylor2025,Hopkins2025}. Conversely, as the object approached the Sun and continued to brighten, compositional studies of its coma began revealing significant differences compared to the chemistry preserved in solar system comets, including enrichments of \ce{CO2}, Ni, Fe, and \ce{CH3OH} \citep{Cordiner2025b,Hutsemekers2026,Roth2026a}, which may point to metal-rich origins. 

However, post-perihelion measurements demonstrated elevated isotopic ratios of $\mathrm{D/H}= (0.98\pm0.06)\%$, $\ce{^12C/^13C}=123-191$, and $\ce{^14N/^15N}=343^{+454}_{-124}$ in water, CO, \ce{CO2}, and CN, respectively \citep{Salazar-Manzano2026,Cordiner2026,Opitom2026}. Studies of isotopic ratios in cometary molecules are particularly powerful tracers of formation conditions, as isotopic fractionation in elements such as H and C is favored in cold environments such as the interstellar medium and the outer protoplanetary disk. On the one hand, the galactic \ce{^13C} abundance continually grows with time owing to stellar nucleosynthesis, including a rapid increase after the first 4 Gyr of the Milky Way from novae production \citep{Botelho2020,Romano2022}. On the other hand, deuteration in primitive environments is initiated by the ion-molecule exchange reaction \citep[e.g.,][]{Millar1989}

\begin{equation}\label{eqn:eqn1}
\ce{H3+ + HD <=> H2D+ + H2} + \Delta E
\end{equation}

\noindent where the endothermic reverse reaction is disfavored when the temperature and ortho-to-para ratio of \ce{H2} are low (see Discussion). Achieving high molecular D/H ratios is further enhanced in the presence of low-metallicity and/or high cosmic ray irradiation rates \citep{Pagani2011,Jensen2021}. Taken together with the dynamical evidence for an origin in a low-metallicity system in the Galactic disk, \cite{Cordiner2026} argued that the high D/H and \ce{^12C/^13C} ratios in 3I/ATLAS require its formation in the interstellar or protostellar phase of an origin system with low metallicity and a high cosmic ray irradiation rate as long as $10-12$ Gyr ago. 

If this were the case, enhanced deuteration may be expected to extend beyond the case of water in 3I/ATLAS. Organic molecules, including methanol and formaldehyde, show D/H ratios up to an order of magnitude higher than that for water toward star-forming regions and protostars in the nearby Milky Way \citep[e.g.,][]{Cazaux2011,Taquet2019}. Thus, methane, representing the simplest alkane and a molecule commonly detected in comets via infrared spectroscopy, may be expected to achieve higher deuteration than water in primitive environments, which may then be passed onto the ices incorporated into cometary nuclei. Unfortunately, there is a severe paucity of \ce{CH3D} detections in comets and in  extrasolar sources owing to intrinsically weak spectral features \citep[e.g.,][]{Riaz2022} and the deuteration of methane in primitive environments is observationally unconstrained. 

Here we report analysis of molecular emission in the coma of the interstellar object 3I/ATLAS measured with the James Webb Space Telescope (JWST) NIRSpec Integral Field Unit (IFU) on UT 2025 December 22 and 23. The observations were obtained at a heliocentric distance (\rh{}) of 2.4 au while the object was at a distance ($\Delta_{\mathrm{JWST}}$) of 1.8 au from the telescope. Additional details of the observations are provided in Methods.  Identifications of detected molecular vibrational bands are given in Supplementary Data. These observations detected ro-vibrational transitions of \ce{CH4} and \ce{CH3D} along with nearby emission from \ce{CH3OH}, \ce{C2H6}, \ce{H2CO}, \ce{H2O}, and CO. To date, \ce{CH3D} has only been detected in a single solar system comet, 67P/Churyumov-Gerasimenko, through in-situ mass spectrometry measurements with the Rosetta spacecraft \citep{Mueller2022,Biver2024b}. The detection of \ce{CH3D} and \ce{CH4} in 3I/ATLAS provides a direct measure of the deuteration in alkanes at the time of planet formation in an extrasolar system. 

\section{Results}\label{sec2}
The object's continuum photocenter was chosen to represent the nucleus position, from which spectra were extracted in a $1''.5$ diameter circular aperture (corresponding to a projected radial distance of $\sim980$ km at the distance of 3I/ATLAS). The key parameters derived in this study are the production rate ($Q$, the number of molecules subliming from the comet nucleus per second) and the rotational temperature (\trot{}, K) for each detected species. These quantities were derived using molecular emission models and the Optimal Estimation Method implemented in the NASA Planetary Spectrum Generator \citep[PSG;][]{Villanueva2018}, which includes a correction for any optical depth effects in the major species \citep[see ][for further details]{Villanueva2025}. 

\begin{figure*}
\begin{center}
\includegraphics[width=\textwidth]{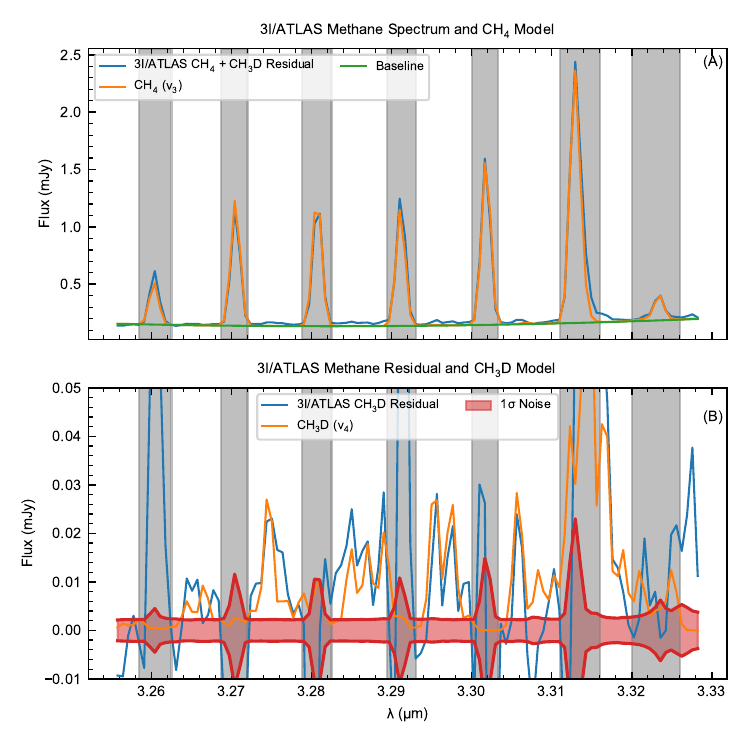}
\end{center}
\caption{\textbf{(A).} 3I/ATLAS \ce{CH4} $+$ \ce{CH3D} residual spectrum, generated by extracting a spectrum in a $1''.5$ diameter aperture centered on the nucleus position and subtracting forward models of \ce{CH3OH}, \ce{H2CO}, \ce{C2H6}, and OH* emission. The best-fit \ce{CH4} model is overplotted. \textbf{(B).} 3I/ATLAS \ce{CH3D} residual spectrum generated by subtracting the best-fit \ce{CH4} model and the spectral baseline from the upper panel. The best-fit \ce{CH3D} model and the $1\sigma$ noise envelope (which includes instrumental noise and uncertainties in the forward molecular models) are also shown. The gray shaded regions denote the positions of \ce{CH4} and blended \ce{CH3D} lines, which were masked when fitting \ce{CH3D} (see Methods).}
\label{fig:ch3d}
\end{figure*}

Isolating the \ce{CH4} and \ce{CH3D} signal requires a careful treatment of any species with potentially blended signatures at the $R\sim2700$ resolving power of the JWST NIRSpec IFU. Among species previously detected in cometary atmospheres with strong transitions near those of methane \citep{Biver2024b}, these include \ce{CH3OH}, \ce{H2CO}, \ce{C2H6}, \ce{NH2}, NH, \ce{CH3CN},  \ce{^13CH4}, and OH* \citep[prompt emission, a vibrationally excited photodissociation product of \ce{H2O} which traces the production rate and spatial distribution of the latter;][]{Bonev2006}. We searched for all of these in the JWST spectra, yet only \ce{CH3OH}, \ce{H2CO}, and \ce{C2H6} were clearly detected. 

\begin{table}
\caption{\label{tab:tab1} Molecular production rates and temperatures in 3I/ATLAS}
\centering
\begin{tabular}{cccccc}
\hline\hline
Molecule & $Q_x$ & \trot{} & $Q_x/Q$(CO) & $Q_x/Q$(\ce{H2O}) & $\langle Q_x/Q(\ce{H2O})\rangle$  \\
   & ($10^{26}$ \ps{}) & (K) & (\%) & (\%) & (\%) \\
\hline
\multicolumn{6}{c}{2025 December 22, G235H/F170LP} \\
\ce{H2O} & $8.70\pm0.06$ & $23.0\pm0.3$ & ... & ... & ... \\
\hline
\multicolumn{6}{c}{2025 December 23, G395H/F290LP} \\
CO & $35.0\pm0.2$ & $42.3\pm0.4$ & 100 & $356\pm5$ & $5.2\pm1.3$  \\
\ce{H2O} & $9.82\pm0.13$ & (23) & $28.0\pm0.4$ & 100 & ...  \\
\ce{CH3OH} & $1.37\pm0.04$ & $34.4\pm0.8$ & $3.92\pm0.09$ & $13.7\pm0.4$ & $2.06\pm0.20$  \\
\ce{H2CO} & $0.039\pm0.009$ & (34) & $0.11\pm0.03$ & $0.40\pm0.09$ & $0.31\pm0.06$ \\
\ce{CH4} & $1.90\pm0.02$ & $41.2\pm0.5$ & $5.43\pm0.06$ &  $19.3\pm0.4$ & $0.78\pm0.09$  \\
\ce{C2H6} & $0.20\pm0.02$ & (41) & $0.58\pm0.06$ & $2.08\pm0.21$ & $0.55\pm0.08$  \\
\ce{CH3D} & $0.25\pm0.03$ & (41) & $0.72\pm0.07$ & $2.57\pm0.25$ & ...  \\
\hline
\end{tabular}
\parbox{10cm}{Note --- $Q_x$ is the molecular production rate of each species. \trot{} is the molecular rotational temperature. Values in parentheses are assumed. $\langle Q_x/Q(\ce{H2O})\rangle$ are mean abundances with respect to \ce{H2O} among comets measured \citep{Biver2024b,DelloRusso2016}.}
\end{table}

We determined $Q$(\ce{CH3OH}) and $Q$(\ce{H2CO}) from modeling of isolated bands elsewhere in the JWST spectra. Despite a lack of clearly detected unblended lines of OH*, we conservatively included it in our models by setting $Q(\ce{OH}^*)=Q(\ce{H2O})$ based on analysis of spectrally isolated \ce{H2O} emission. We then used these quantities to generate forward models of emission for \ce{CH3OH}, \ce{H2CO}, \ce{C2H6}, and OH* expected near the position of the methane emission and subtracted them to generate a \ce{CH4} and \ce{CH3D} residual spectrum of 3I/ATLAS. Figure~\ref{fig:ch3d} shows the resulting spectrum. Our retrieved production rates are summarized in Table~\ref{tab:tab1}. 

Table~\ref{tab:tab1} shows that CO was the most abundant molecule in 3I/ATLAS's coma during our observations. Thus, we primarily calculated molecular abundances with respect to CO for all species in this study. \ce{H2O} is often the dominant coma molecule in comets measured at \rh{}$<3$ au and is the standard compositional reference molecule in other comets \citep{Biver2024b,DelloRusso2016}, so we computed molecular abundances with respect to \ce{H2O} for ease of comparison. Our derived D/H ratio for methane is $(3.33\pm0.31)\%$. 

In terms of overall composition, our results are consistent with other studies which indicate 3I/ATLAS was in general enriched in volatiles with respect to water, with the exception of sulfur-bearing species \citep[e.g.,][]{Cordiner2026,Salazar-Manzano2026,Belyakov2025,Biver2026}. However, post-perihelion measurements at \rh{}$>2$ au must be interpreted in the context of the \ce{H2O} ice line: as comets cross \rh{}$\sim2-3$ au, \ce{H2O} sublimation becomes less vigorous with attenuating solar insolation, leaving CO and \ce{CO2} to become primary activity drivers. Since the majority of comet composition studies to date were undertaken interior to \rh{}$\sim2$ au in \ce{H2O}-dominated comae \citep[e.g.,][ and references therein]{Biver2024b,DelloRusso2016}, this caveat must be kept in mind. However, methane is among the most volatile molecules routinely measured in comets and should certainly be fully active at the \rh{} = 2.4 au of our study \citep{DelloRusso2016}, so our derived D/H ratio should not be affected by the transition from \ce{H2O}- to CO-dominated outgassing in 3I/ATLAS. 

\section{Discussion}\label{sec3}
Of the trace species detected, the most notable is \ce{CH3D}. It has proven extremely difficult to measure in solar system comets even with the most powerful ground- and space-based facilities. Aside from $3\sigma$ upper limits provided in a few bright comets \citep[D/H$<(0.5-1)\%$;][]{Gibb2012,Bonev2009,Kawakita2005}, the only secure detection of \ce{CH3D} in a comet was through in-situ mass spectroscopy of comet 67P/Churyumov-Gerasimenko (itself one of the most D-enriched solar system comets found to date) performed by the Rosetta spacecraft \citep{Mueller2022,Biver2024b}. Thus, this JWST study represents not only its rare detection in an interstellar object, but through remote sensing of a cometary atmosphere in general.

Aside from comet 67P/Churyumov-Gerasimenko, the D/H ratio for methane in the solar system has been measured for Venus, the giant planets, Titan, and the dwarf planets Eris and Makemake. The D/H ratio for organics has also been measured in interplanetary dust particles (IDPs), carbonaceous meteorites, and ultra-carbonaceous Antarctic micrometeorites (UCAMMs; see Extended Data Table~\ref{tab:dh}). The value in 3I/ATLAS is at least a factor of 1.4 higher than any other direct measure for organic molecules in the solar system, with the closest being D/H = $(1.5\pm0.5)\%$ for organic radicals in carbonaceous chondrite meteorite Orgueil \citep{Gourier2008}. 

Our D/H ratio of $(3.33\pm0.31)$\% for methane is higher than that found for comet 67P/Churyumov-Gerasimenko, $(0.241\pm0.029)\%$, by a factor of $14\pm2$. Yet the relative levels of deuteration measured for water and methane in the two comets provide greater insight. Compared to the D/H ratio in water reported by \cite{Mueller2022}, the D/H ratio in methane measured in 67P/Churyumov-Gerasimenko by Rosetta was higher by a factor of $4.8\pm0.7$. A  significantly lower D/H ratio for water was later reported by analyzing post-perihelion measurements taken exclusively at nucleocentric distances $>120$ km, in which case the D/H ratio in methane was instead higher by a factor of $9.3\pm1.7$ \citep{Mandt2024}. The D/H ratio for methane in 3I/ATLAS is higher than the same ratio in water reported by \cite{Cordiner2026} by a factor of $3.4\pm0.4$. Although the sample for comets is very small ($n=2$), this is consistent with a trend of higher deuteration in organic molecules compared to water observed in other solar system bodies, as well as toward extrasolar sources \citep{Ceccarelli2014,Cleeves2016}.

\begin{figure*}
\begin{center}
\includegraphics[width=\textwidth]{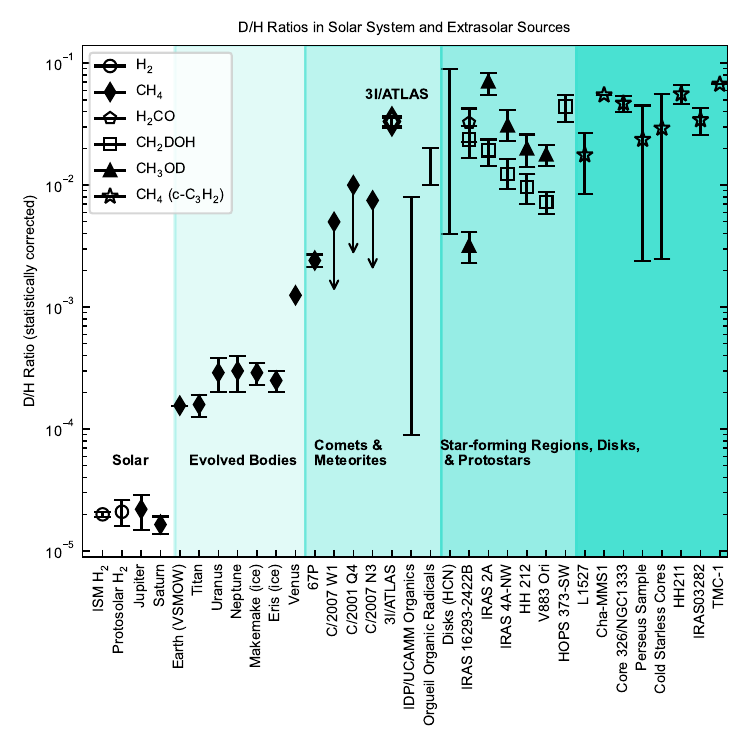}
\end{center}
\caption{D/H ratios in solar system and extrasolar sources. Ratios are for methane except when noted otherwise. Sources for which the D/H ratio in $c-$\ce{C3H2} was used to estimate the D/H in \ce{CH4} are denoted with a $\star$. References and values for each object are given in Extended Data Table~\ref{tab:dh}.}
\label{fig:dh}
\end{figure*}

Beyond the solar system, \ce{CH3D} has been detected toward three proto-brown dwarfs \citep{Riaz2022}, yet secure measurements can be found almost nowhere else. However, \ce{CH4} was not directly measured in these sources (it was inferred from \ce{CH3D} and the \ce{DCO^+}/\ce{HCO^+} ratio), so the D/H ratio in methane is observationally unconstrained beyond the heliopause to date. Owing to this paucity of direct measures, we estimated the CH$_3$D/CH$_4$ isotopic ratios in several sources using deuteration in cyclopropenylidene (${\it c-} \rm C_3H_2$), a small, cyclic molecule that is ubiquitous in the neutral interstellar medium (see Methods). Our results suggest that the D/H ratios in methane in these environments may reach $\geq4-7\%$. Figure~\ref{fig:dh} shows a comparison of D/H ratios for methane and other organic molecules in solar system and extrasolar sources.  For direct measures in extrasolar sources, the D/H ratio for methane in 3I/ATLAS is most similar to those for other organic molecules, such as formaldehyde and methanol, in star-forming regions and protostars. 

In the absence of direct observational constraints on extrasolar methane D/H ratios, astrochemical models may provide insights. We note that to the best of our knowledge, the D/H ratios of icy species under low metallicity environments with high cosmic ray irradiation rates \citep[such as the proposed origin system for 3I/ATLAS;][]{Cordiner2026} have not yet been investigated by astrochemical models. Thus, we begin by discussing deuteration chemistry in environments consistent with the local Milky Way. In cold ($\sim10$ K) dense matter, cosmic rays ionize molecular hydrogen, leading to ion-molecule reactions that produce \ce{H3+}. In turn, \ce{H3+} can react with HD to form \ce{H2D+} (Reaction~\ref{eqn:eqn1}) where the  rate of the endothermic reverse process depends on the ortho/para spin ratio of the  hydrogen molecules (OPR=${\it o-} \rm H_2 $/${\it p-}\rm H_2 $);  destruction of H$_2$D$^+$ is accelerated even at low temperatures by reaction with molecules in the higher-energy ${\it o-} \rm H_2 $ state. Gas-phase ion-molecule reactions involving H$_2$D$^+$ subsequently lead to the incorporation of D atoms into interstellar molecules. 

Late in the evolution of  prestellar cores, significant depletion of CO molecules removes a major destruction pathway for molecular ions, allowing \ce{D2H+} and \ce{D3+} to form efficiently via \ce{H2D+ + HD <=> D2H+ + H2} and \ce{D2H+ + HD <=> D3+ + H2} \citep{Roberts2003}. In this case, dissociative electron recombination reactions release more D atoms into the gas, elevating the atomic D/H ratio and  leading to grain-surface chemistry  producing high D/H ratios ($\sim 10^2-10^4 \times$ the cosmic ratio) and multiply-deuterated molecules \citep[e.g., \ce{ND3}, \ce{D2CO};][]{Ceccarelli2014}.

On grain surfaces, accreted carbon atoms provide the starting point for the formation of organic molecules and their deuterated isotopologues through a series of H and D atom addition reactions, and so we may expect these fractionation characteristics to also be evident. Models of deuterium fractionation in molecular clouds should therefore be required to self-consistently calculate the time-evolution of the ortho and para spin states of H$_3^+$ and H$_2$, as well as those of H$_2$D$^+$, D$_2$H$^+$, and D$_3^+$ \citep{Hugo2009,Sipila210}. A further consequence of the time-dependence  of H$_2$  spin conversion and CO depletion is that the ice mantle structure develops a  D/H fractionation gradient: molecular  D/H ratios are  highest in the (late-forming) monolayers closest to the surface \citep{Taquet2014}. 

Theoretical models of gas-grain deuterium fractionation in  prestellar core and protostar formation that include all  of these processes and assume  metallicity values consistent with the local Milky Way  \citep{Taquet2014,Furuya2016,Jensen2021} can produce the  range of HDO/H$_2$O ratios found in interstellar ice mantles, hot corinos, and Solar System comets, as well as the  ratio recently reported for 3I/ATLAS \citep[$\approx$ 1\%;][]{Cordiner2026}. Several can reproduce the observed range of deuteration in organic molecules, such as methanol and formaldehyde, observed toward dark clouds and protostars, although these models do not include consideration of molecular OPR \citep[and in some cases, grain mantle structure;][]{Aikawa2012,Albertsson2013,Taquet2019,Jorgensen2018}. Unfortunately, due to the absence of  observed \ce{CH3D/CH4} ratios, few studies actually list the computed gas and ice phase ratios. Of the few that do, including models of deuteration processes in the protoplanetary disk \citep[e.g.,][]{Cleeves2016}, none predict $\mathrm{D/H}>1\%$. On the other hand, these models predict that differences in formation mechanism (gas-phase vs.\ grain-surface chemistry) should result in lower D/H ratios in methanol compared to methane \citep{Cleeves2016}. This is qualitatively consistent with IRAM 30-m measurements of 3I/ATLAS near perihelion, which found a $3\sigma$ upper limit on $\ce{CH3OD}/\ce{CH3OH}<5.3\%$ \citep{Biver2026}. This upper limit rules out methanol deuteration in significant excess of our methane D/H ratio in 3I/ATLAS.

What mechanisms may then explain the high D/H ratio in methane measured in 3I/ATLAS? As earlier noted, the D/H ratio for methanol exceeds that of water by a factor of $\sim10$ in star-forming regions and protostars \citep[e.g.,][]{Taquet2019}. This has been interpreted as reflecting the difference in the formation times of water and methanol on grain surfaces in the prestellar phase. Water and methanol are formed on grain surfaces through successive H/D addition reactions to O and CO, respectively. The atomic D/H ratio in the gas determines the relative deuteration available for these grain-surface reactions at the formation time of each molecule. Since the atomic D/H ratio in the gas increases with time at low temperatures (such as encountered in the prestellar phase), the higher D/H ratios observed for methanol suggest that it formed later than water \citep{Cazaux2011,Taquet2012,Ceccarelli2014}.  The high D/H ratios in methane compared to water in 3I/ATLAS and 67P/Churyumov-Gerasimenko may therefore be evidence of a relatively late molecular formation time in the prestellar cloud. Indeed, methane can be efficiently formed on grains by the H/D addition reactions to C \citep{Qasim2020}.

Given the difficulties in achieving the D/H measured for methane in 3I/ATLAS using models that assume conditions consistent with the local Milky Way, we discuss the impact that low temperatures, low metallicity, and high cosmic ray irradiation rates would have on the deuteration of organic molecules in the proposed origin system of 3I/ATLAS \cite{Cordiner2026}. Sustained low temperatures maintain the production of increasingly deuterated ions along the \ce{H3+ ->[HD] H2D+ ->[HD] D2H+ ->[HD] D3+} ladder and prevent the endothermic reverse reactions from proceeding, thereby providing feedstock for deuterated organic molecules derived from both gas-phase and grain-surface chemistry. Furthermore, increased deuteration can be passed on to methane through both the \ce{H3+} $\buildrel \mathrm{HD} \over \longrightarrow...\buildrel \mathrm{HD} \over \longrightarrow$\ce{D3+}  ladder as well as through \citep{Millar1989}

\begin{equation}\label{eqn:eqn2}
\ce{CH3+ + HD <=> CH2D+ + H2} + \Delta E
\end{equation}

whereas water deuteration experiences contributions from only the former. Thus, the D/H ratios for methane and organic molecules can continue to increase even as temperatures rise beyond $\sim30$ K (such as in the protoplanetary disk stage) given the higher exothermicity of Reaction~\ref{eqn:eqn2} \citep{Aikawa1999,Cleeves2016}.

Similarly, low metallicity environments help to preserve deuterated ions by reducing the amount of gas-phase CO available to destroy \ce{H3+} and other molecular ions \citep{Pagani2011}. Under solar metallicity conditions, such depletion does not occur until CO freeze-out in the late prestellar stage; thus, low metallicity increases the time available for ion-molecule exchange reactions to efficiently increase the atomic D/H ratio in the gas. Finally, high cosmic ray irradiation rates increase the abundance of \ce{H3+} (as well as \ce{H+}), thereby driving up deuteration by multiple avenues \citep{Jensen2021}. On the one hand, additional \ce{H3+} increases the reactants available to drive production of \ce{H2D+} via Reaction~\ref{eqn:eqn1}. On the other hand, the extra availability of ionized hydrogen enables faster conversion of $o-$\ce{H2} to $p-$\ce{H2} via proton-exchange reactions, thus decreasing the OPR and the efficiency by which the \ce{H3+} $\buildrel \mathrm{HD} \over \longrightarrow...\buildrel \mathrm{HD} \over \longrightarrow$\ce{D3+} ladder can proceed in reverse at low temperatures. Higher ionization rates (up to $10\times$ the local Galactic value) may be expected in the low metallicity environment required to explain the \ce{^12C/^13C} ratio in 3I/ATLAS \citep{Cordiner2026}, where reduced dust abundances would less efficiently attenuate interstellar radiation fields. Collectively, these observational and model results help to frame the methane D/H ratio in 3I/ATLAS as an expected consequence of the conditions required to produce the measured isotopic ratios in its water, CO, and \ce{CO2}.

\clearpage

\setcounter{figure}{0}
\setcounter{table}{0}
\renewcommand{\figurename}{Extended Data Figure} 
\renewcommand{\tablename}{Extended Data Table} 

\section*{Methods}\label{howto}

\subsection*{JWST Observations}\label{obs}
Observations of 3I/ATLAS were carried out using the JWST NIRSpec IFU \citep{Boker2022} on UT 2025 December 22 using the G235H/F1270LP grating (covering $\lambda=1.66 - 3.05$ $\mu$m) and December 23 using the G395H/F290LP grating (covering $\lambda=2.87-5.14$ $\mu$m) under General Observer program 5094. These gratings provide a spectral resolving power $\lambda/\Delta\lambda\sim2700$. The comet's observing geometry varied from \rh{}$ = 2.37 - 2.42$ au and $\Delta_\mathrm{JWST} = 1.79 - 1.80$ au, with a solar phase angle $\phi=22.7^{\circ}-21.6^{\circ}$ and was tracked using JPL Horizons ephemeris orbit solution \#42. A campaign to obtain accurate astrometry for and refine the ephemeris of 3I/ATLAS is detailed in \cite{Cordiner2026}.

The G235H grating was executed with a single 642 s exposure beginning UT 2025-12-22 03:36, followed by $5\times700$ s exposures with G395H  starting UT 2025-12-23 08:07. All observations used a four-point dither pattern. Background exposures with identical circumstances (exposure time, grating settings) were planned offset from the comet position by $300''$; however, two of the five background exposures for the G395H grating failed owing to background star issues, with follow-up attempts scheduled for later in 2026. Careful examination of the available background exposures did not reveal emission from any known interstellar infrared sources or zodiacal light, so analysis proceeded without background subtraction to maximize SNR. Exposures were processed using the JWST Pipeline version 1.20.2 with CRDS jwst\_1464.pmap context files and aligned onto a common spatial-spectral axis using the Drizzle algorithm \citep{Bushouse2025}.

\subsection*{Spectral Modeling}\label{modeling}
\cite{Cordiner2026} reported analysis of ALMA CO and HCN measurements in 3I/ATLAS carried out on 2025-Dec-22 between UT 07:32--08:36, as well as analysis of \ce{H2O}, CO, and \ce{CO2} (along with their isotopologues) from the same JWST observations reported here. We adopted their relevant quantities (e.g., HDO/\ce{H2O}, gas expansion speeds) for the purposes of this analysis. Supplementary Table~\ref{tab:bands} provides a summary of the detected species and their ro-vibrational emission bands in this work. 

We extracted and modeled spectra from a nucleus-centered $1''.5$ diameter aperture. Although the very central 1--2 spaxels are affected by the instrumental point-spread function (PSF), opacity effects are minimal within such an aperture at the low molecular production rates measured in this study. We used the \texttt{jwstComet} package \citep{Roth2026b}, which provides for flexible spectral extraction from JWST IFU data cubes using functions from the \texttt{astropy} and \texttt{photutils} libraries, followed by automated interfacing with the NASA PSG API. The data cubes were converted from units of MJy sr$^{-1}$ to Jy pixel$^{-1}$, then spectra were extracted from the data cubes using the \texttt{photutils} CircularAperture function and summed within the aperture using the aperture\_photometry function (method = `subpixel', subpixels = 10). Fluxes and $1\sigma$ noise on a per-spaxel basis were derived from the \texttt{SCI} and \texttt{ERR} extensions of the FITS files. 

In the G395H grating, CO, \ce{CH3OH}, \ce{C2H6}, \ce{H2CO}, \ce{CH4}, \ce{CH3D}, and \ce{H2O} were sampled simultaneously, while the G235H setting measured \ce{H2O}. In G395H, CO has strong emissions near 4.5 $\mu$m, whereas \ce{H2O} has hot band emission throughout the $4-5$ $\mu$m region. Although this \ce{H2O} emission was sufficiently strong for measuring $Q$(\ce{H2O}), we could not retrieve a well-constrained rotational temperature. Instead the \ce{H2O} \trot{} was determined from the G235H setting, then fixed when analyzing \ce{H2O} emission (and forward modeling OH*) in G395H.

We first worked to determine $Q$(CO) (relevant for setting the activity level in PSG) and $Q$(\ce{H2O}) (necessary for generating forward models of OH*) in the aforementioned aperture. Uncertainties on the PSG retrieved parameters were obtained from the diagonal elements of the covariance matrix, scaled by the square root of the reduced $\chi^2$ statistic, which includes uncertainties in the spectral baseline of the fit. All spectra were baseline subtracted using second or third-order polynomial fits. We used the lowest order polynomial baseline capable of reproducing the spectral shape and avoided higher-order polynomials to prevent introducing spurious features into the spectra. In all instances, we fit the spectral baseline simultaneously with the molecular emission, thereby incorporating uncertainties in the baseline fit into the uncertainties on each retrieved quantity (i.e., $Q$ or \trot{}). This baseline accounts for continuum emission from the dust and nucleus, as well as scattered sunlight and instrumental artifacts. Techniques employing simultaneous fitting of the continuum baseline and molecular emission models have been applied to decades of cometary infrared spectroscopy studies \citep[e.g.,][]{DiSanti2003,Bonev2014,DiSanti2017,Faggi2019,Ejeta2024}. We used a fixed spectral resolution element based on curves for dispersion as a function of wavelength provided by the Space Telescope Science Institute (\url{https://jwst-docs.stsci.edu/jwst-near-infrared-spectrograph/nirspec-instrumentation/nirspec-dispersers-and-filters\#gsc.tab=0}).

As shown in Table~\ref{tab:tab1}, there are significant differences in \trot{} between trace species, especially the apolar and polar species. However, the PSG only allows specification of a single \trot{} during a retrieval; thus, it is not possible to model CO (\trot{} = 42 K) and \ce{H2O} (\trot{} = 23 K) in 3I/ATLAS during a single PSG simulation. This difficulty is most acute when modeling species whose strongest transitions are in spectrally crowded regions, such as \ce{CH4}, \ce{CH3D}, and \ce{C2H6}. We therefore retrieved relevant quantities ($Q$ or \trot{}) for potentially confusing species (e.g., \ce{CH3OH}) from spectrally isolated bands whenever possible, then generated forward models of their emission to subtract away from species that are only available in spectrally crowded regions. We detail each of these instances below.

Following \cite{Cordiner2026}, we set the gas expansion speed, $v_\mathrm{exp}$, to 0.345 \kms{} for \ce{CO} and to 0.310 \kms{} for all other species based on velocity resolved ALMA measurements of CO and HCN. We first retrieved $Q(\ce{H2O})$ and \trot{}(\ce{H2O}) from the G235H setting on December 22. To determine $Q(\ce{H2O})$ from the G395H setting on December 23, we masked the strong nearby CO and OCS emission, fixed \trot{}(\ce{H2O}) = 23 K, and analyzed \ce{H2O} emission between $4.45 - 5.1$ $\mu$m. Next we analyzed the CO emission, retrieving $Q$(CO) while allowing \trot{}(CO) to freely vary. The retrieved values, including molecular production rates and rotational temperatures, are given in Table~\ref{tab:tab1}. Owing to a lack of spectrally isolated OH* emission lines clearly detected in our spectral regions of interest at the resolving power of JWST, we generated forward models of OH* by setting $Q(\mathrm{OH}^*)=Q(\ce{H2O})$ in all subsequent steps. 

Next, we retrieved $Q$(\ce{CH3OH}) and \trot{}(\ce{CH3OH}) by fitting the $\nu_3$ vibrational band near 3.5 $\mu$m simultaneously with $Q$(\ce{H2CO}) from the $\nu_1$ and $\nu_5$ bands near 3.6 $\mu$m. Aside from several potentially blended OH* lines, these are spectrally isolated bands of each species which could be used to generate forward models when analyzing \ce{CH4}, \ce{C2H6}, and \ce{CH3D}.

Owing to the proximity of the \ce{C2H6} $\nu_5$ vibrational band to confusing emission from \ce{CH3OH}, \ce{H2CO}, and \ce{CH4}, we took an iterative approach to its retrieval. The \ce{C2H6} $\nu_5$ band does not emit appreciably at wavelengths shorter than 3.32 $\mu$m, so we first worked to measure $Q(\ce{CH4})$ by analyzing emission from $3.20 - 3.32$ $\mu$m. We generated forward models of \ce{CH3OH} ($\nu_2,\nu_3,\nu_9$), \ce{H2CO} ($\nu_1+\nu_6$), and OH* spectra at their respective rotational temperatures (34 K for \ce{CH3OH} and \ce{H2CO} and 23 K for OH*) and previously determined $Q$'s and subtracted them from our observed 3I/ATLAS spectrum. 

We detected the \ce{CH4} $R(0)$ through $R(8)$ transitions, the $Q$-branch, and several $P$-branch lines (although the latter becoming increasingly blended with \ce{CH3OH} and \ce{C2H6}). Similar to challenges fitting \ce{CO2} faced by \cite{Cordiner2026}, we found that the higher-$J$ \ce{CH4} $R$-branch lines were not fit well by our radiative transfer models, likely owing to non-LTE effects at the very low temperatures and production rates in 3I/ATLAS during our observations. Namely, the higher-$J$ lines depart from approximately LTE conditions more quickly than their lower-$J$ counterparts. In turn, the PSG-produced baseline underneath the higher-$J$ lines was skewed towards negative values. To test to what extent this may affect our fit, we fit a second order polynomial to the \ce{CH4} extract before modeling the molecular line emission (Supplementary Figure~\ref{fig:baseline}). This model-independent baseline did not suffer from the same challenges near the higher-$J$ lines, yet it clearly significantly overfits the baseline near the critically important trace emission between the towering \ce{CH4} lines (especially the lower-$J$ lines). Nevertheless, it is informative that we retrieve $Q(\ce{CH4})=(1.99\pm0.03)\times10^{26}$ \ps{} and \trot{}(\ce{CH4}) = $43.6\pm0.6$ K.

We then returned to fitting the spectral baseline simultaneously with the \ce{CH4} emission in the PSG, yet excluding the higher-$J$ lines ($R(5)-R(8)$) from the fit (i.e., considering extracts with $\lambda>3.255$ $\mu$m). This provided a significantly better baseline fit to the emission between the \ce{CH4} lines and to the lower$-J$ \ce{CH4} lines themselves. Our retrieved values (Table~\ref{tab:tab1}) agree within $2\sigma$ with the results from fitting all $R$-branch lines. Satisfied that restricting the number of \ce{CH4} lines in our fit did not significantly affect the results, we proceeded to work to retrieve $Q(\ce{C2H6})$. We extended the spectral extract to cover wavelength ranges $\lambda=3.255-3.38$ $\mu$m. This adequately covers the strong lines of the \ce{C2H6} $\nu_5$ band but does not yet extend into the strongest features of the \ce{CH3OH} $\nu_9$ band or other potential underlying solid-state features \citep[e.g., polycyclic aromatic hydrocarbons;][]{Woodward2025} at longer wavelengths. We fixed $Q$(\ce{CH4}) and \trot{}(\ce{C2H6}) = \trot{}(\ce{CH4}) and retrieved $Q(\ce{C2H6})$, assuming that it followed the \trot{} of the other apolar species.

Finally, we turned to fitting \ce{CH3D}, extracting a spectrum from $\lambda=3.255 - 3.328$ $\mu$m and then subtracting the forward models of emission for \ce{CH3OH}, \ce{C2H6}, \ce{H2CO}, \ce{CH4}, and OH* (Supplementary Figure~\ref{fig:fits3}) to generate a \ce{CH3D} residual. Since line flux is linearly proportional to molecular production rate, the uncertainty ($\sigma_F)$ in the forward modeled flux for species $i$ ($F_i$) at a wavelength $\lambda$ given a production rate ($Q_i$) and its uncertainty (${\sigma_Q}_i$) is ${\sigma_F}_i (\lambda)=F_i (\lambda)({\sigma_Q}_i /Q_i)$. We added the model uncertainties for each subtracted forward model in quadrature to the $1\sigma$ instrumental uncertainties on the \ce{CH3D} residual spectrum before retrieving $Q(\ce{CH3D})$. As shown in Figure~\ref{fig:ch3d}, the \ce{CH4} line fluxes are up to two orders of magnitude higher than the \ce{CH3D} line fluxes, yet the \ce{CH3D} features are clearly detected above the instrumental noise envelope. Given the dynamic range between the \ce{CH4} and \ce{CH3D} lines, even slight mismatches between the observed and modeled \ce{CH4} lines produce residuals that are comparably strong to the \ce{CH3D} lines themselves. Thus, we masked the positions of the \ce{CH4} lines while retrieving \ce{CH3D} to prevent uncertainties in the \ce{CH4} fit from propagating into the \ce{CH3D} retrieval. Furthermore, this masking removed strongly blended \ce{CH3D} features from the analysis, such as that near the 3.315 $\mu$m \ce{CH4} $Q$-branch. We then fixed \trot{}(\ce{CH3D}) = \trot{}(\ce{CH4}) and allowed $Q(\ce{CH3D})$ to vary as a free parameter. We tested the effects of varying the order of the polynomial spectral baseline from two to four, finding consistency regardless of our choice. Supplementary Table~\ref{tab:baselines} provides a comparison of the derived molecular production rates, \trot{}, and D/H ratio for methane as a function of polynomial baseline order. Our overall modeling process is shown in Supplementary Figure~\ref{fig:fits2} and Supplementary Figure~\ref{fig:fits3}, including a total molecular emission model for the 3.3 $\mu$m region in 3I/ATLAS.

\subsection*{Estimating Methane D/H Ratios using $c-$\ce{C3H2}}\label{cyclic}
Here we detail our motivation and formalism for estimating the D/H ratio in \ce{CH3D} in primitive environments using the measured D/H ratio in hydrocarbons as a proxy.  We discuss interstellar inheritance of pristine material that was fractionated during the star  formation process as well as isotope chemistry in protoplanetary disks. We also address the viability of estimating the D/H ratio in methane through interstellar chemistry by examining observations of prestellar and protostellar cores. 

Calculations and observations of water deuteration support the view that cometary matter was inherited from the prestellar/protostellar phase of Solar System formation and not produced in situ \citep[e.g.,][]{Furuya2017,Tobin2023}. Additional support for this viewpoint comes from the high \ce{D2O}/HDO ratio measured in comet 67P/Churyumov-Gerasimenko by Rosetta \citep{Altwegg2017} as well as studies of protoplanetary disk chemistry. If cosmic rays were ineffective for ionizing the protosolar nebula \citep{Dolginov1994,Cleeves2013},  ion-molecule chemistry would have been shut down  and  the water D/H ratios in comets and Earth's oceans could not have been produced \citep{Cleeves2014}. For CH$_4$ deuterium fractionation, calculations  which include  cosmic-ray ionization rates  below that commonly assumed for disks \citep[e.g.,][]{Nomura2023}, but where X-rays  can also provide a source of ionization,  only produce  D/H for methane of $\approx $ 0.2-0.6\% \citep{Cleeves2016}. Hence, comparison with  the  D/H ratio in 3I/ATLAS of $(3.33\pm0.31)\%$ for methane  suggests that, like water,  its ices were inherited and not formed in its natal disk.

Observations of CH$_3$D/CH$_4$ isotopic ratios do not exist for star-forming  environments due to the difficulty in detecting gaseous CH$_3$D.  One exception is the tentative detection of CH$_3$D emission toward the protostar IRAS 04368+2557  in  the  L1527 molecular cloud \citep{Sakai2012}.  This object belongs to a population of protostellar cores that exhibits a so-called warm  carbon-chain chemistry \citep[WCCC;][]{Sakai2013} comprising of carbon-chain molecules and simple hydrocarbons usually found in cold molecular clouds. In such `lukewarm corinos', elevated dust temperatures in the protostellar envelope, $T \sim$ 30 K,  allow methane (and CO) to evaporate from icy grain mantles and subsequently drive the observed gas-phase hydrocarbon chemistry,  e.g. forming cyanoacetylene in the sequence 

\begin{equation}\label{CH4seq}
 {\rm{CH_4   \buildrel C^+ \over \longrightarrow C_2H_3^+ \buildrel e^- \over \longrightarrow C_2H_2 
\buildrel CN  \over \longrightarrow HC_3N   }  } 
\end{equation}

Although CH$_4$ has recently been detected in the L1527 ice mantles \citep{Devaraj2026}, CH$_3$D has not. This, and the difficulty in detecting gaseous CH$_4$ in cool material, as opposed to hot cores \citep{Sakai2012}, means that it is not possible to derive the CH$_3$D/CH$_4$ isotopic ratio in L1527, or indeed in any WCCC source. However, following \cite{Sakai2012},  we can use hydrocarbon (i.e., $c-$\ce{C3H2}) D/H ratios measured as a proxy for that in methane. 

Cyclopropenylidene (${\it c-} \rm C_3H_2$) is small, cyclic molecule that is ubiquitous in the interstellar medium. Both it and its mono-deuterated form can be produced from CH$_4$ and CH$_3$D, e.g. through 

\begin{equation}\label{CH3Dseq}
 {\rm{CH_3D   \buildrel C^+ \over \longrightarrow C_2H_2D^+ \buildrel e^- \over \longrightarrow C_2HD
\buildrel C^+  \over \longrightarrow C_3D^+ \buildrel H_2 \over 
\longrightarrow~{\it c}-C_3H_2D^+  \buildrel e^- \over \longrightarrow  {\it c}-C_3HD   }  } 
\end{equation}
\citep{Millar2024}. Defining $R$(XD) as the ratio of observed column densities  $N$(XD)/$N$(XH), we can estimate the D/H ratio for \ce{CH3D} as $R$(${\it c-} \rm C_3HD$)/4. This estimate is based on the assumption that the evaporated CH$_3$D/CH$_4$  ratio is unaltered  in the gas.

While deuteration by {\color{black}  Reaction (\ref{eqn:eqn1}}) is suppressed at  $T$ \gtsim 30 K, it can be maintained by Reaction~\ref{eqn:eqn2}, which can then deuterate methane  through  

\begin{equation}\label{He+seq}
 {\rm{CH_4   \buildrel He^+ \over \longrightarrow CH_3^+ \buildrel HD \over \longrightarrow CH_2D^+
\buildrel H_2  \over \longrightarrow CH_4D^+ \buildrel CO, e^- \over \longrightarrow ~CH_3D  }  }
\end{equation}

\noindent However, hot core calculations do show that the D/H ratios of ice molecules are unchanged on post-evaporation time-scales of $\sim 10^5$ years \citep{Rodgers1996}.

As each of the two dissociative electron recombination steps in  {\color{black}  sequence (\ref{CH3Dseq}}) has an associated product channel which removes a D atom from the sequence, one has  $R$(${\it c-} \rm C_3HD$) $<$ $R$(${\rm CH_3D}$) with the reduction factor determined by the details of the product branching ratios. Hence, the estimates of the D/H ratio for methane made here should be considered as lower limits. From the $R$(${\it c-} \rm C_3HD$) values observed towards protostars we estimate D/H ratios for methane of 1.7\% in L1527 \citep{Sakai2009}, 5.5\% in Cha-MMS1 \citep{Lis2025}, and $\approx$ 0.25-4.5\% in a sample of 6 cores in Perseus \citep{FerrerAsensio2026}. 

Methane molecules formed on grain surfaces can also be continuously released into the gas by {\it reactive desorption} whereby the energy released in exothermic reactions (such as hydrogen atom addition) can overcome the surface binding energy of the product molecule \citep{Brown1991,Cuppen2024}.  Thus, one can make an  estimate of methane D/H ratios in cold ($\sim$ 10 K) starless/pre-stellar cores. Surveys  of ${\it c-} \rm C_3HD$  towards 27 such sources \citep{Chantzos2018,FerrerAsensio2026} yield D/H $\approx$ 0.25-5.6\%. Only 4 sources match or exceed the ratio in 3I/ATLAS of D/H $=(3.33\pm0.31)$\%: two of the 8 protostellar cores, Cha-MMS1 and Cores 326 in NGC1333, and two of the 27 cold cores, HH211 and IRAS03282. This suggests that D/H ratios for methane $\approx$ 4.5-5.5\% are possible in pre-cometary matter and in fact, based on the $R$(${\it c-} \rm C_3HD$) found in TMC-1 \citep{Gratier2016}, could be as high as 7\%.  As these estimates are lower limits, even larger methane D/H  ratios could be possible. Extended Data Table~\ref{tab:dh} provides values and references for D/H ratios towards sources detailed in Figure~\ref{fig:dh}, including those estimated here.

To conclude, it is uncertain whether theoretical models of either star formation or disk chemistry can reproduce the methane D/H ratio in 3I/ATLAS's ices. This should ideally be confirmed by new chemical models. On the other hand, there is tentative evidence that interstellar/protostellar D/H ratios could match that of 3I/ATLAS.

\section*{Acknowledgments}
\textbf{Correspondence and requests for materials} should be sent to Nathan Roth.

\vspace{12pt}

\noindent This work is based on observations made with the NASA/ESA/CSA James Webb Space Telescope. The data were obtained from the Mikulski Archive for Space Telescopes at the Space Telescope Science Institute, which is operated by the Association of Universities for Research in Astronomy, Inc., under NASA contract NAS 5-03127 for JWST. We gratefully acknowledge the assistance of optical observers who submitted astrometric observations of 3I/ATLAS in the weeks leading up to our observations, to help refine the ephemeris position. In particular, we thank C. Chandler, J. Chatelain, M. Frissell, E. Gomez, S. Greenstreet, W. Hoogendam, C. Holt, K. Meech, H. W. Lin, T. Lister, L. Salazar Manzano, T. Santana-Ros, D. Seligman, D. Singh, Q. Ye, Q. Zhang. Supporting astrometric observations were obtained by the Comet Chasers school outreach program (https://www.cometchasers.org/ ), led by Helen Usher, which is funded by the UK Science and Technology Facilities Council (via the DeepSpace2DeepImpact Project), the Open University and Cardiff University. It accesses the LCOGT telescopes through the Schools Observatory/Faulkes Telescope Project (TSO2025A-00 DFET-The Schools’ Observatory), which is partly funded by the Dill Faulkes Educational Trust, and through the LCO Global Sky Partners Programme (LCOEPO2023B-013). Observers included individuals and representatives from the following schools and clubs: E. Maciulis, A. Bankole, J. Bower, O. Roberts, participants on the British Astronomical Associations’ Work Experience project 2025 from The Coopers Company \& Coborn School; Upminster, UK;  St Marys Catholic Primary School, Bridgend, UK; J. M. Perez Redondo \& Students: A. Matea, L. Guillamet, A. Montoy, and A. Martin from Institut d’Alcarràs, Catalonia, Spain; Louis Cruis Astronomy Club, Brazil; Jelkovec High School, Zagreb, Croatia, and C. Wells at a British Astronomical Association event. This research has made use of NASA’s Astrophysics Data System Bibliographic Services. This research has made use of data and/or services provided by the International Astronomical Union's Minor Planet Center. 

\begin{itemize}
\item \textbf{Data availability:}
All JWST data are available through the Mikulski Archive for Space Telescopes at the Space Telescope Science Institute under proposal ID \#5094 (https://doi.org/10.17909/1jvn-1z72). 

\item \textbf{Code availability:}
The JWST Calibration Pipeline \citep{Bushouse2025} is available from https://github.com/spacetelescope/jwst. The Planetary Spectrum Generator \citep{Villanueva2025,Villanueva2018}, used for modeling cometary infrared emission lines,  is available at https://psg.gsfc.nasa.gov/. The \texttt{jwstComet} software \citep{Roth2026b}, used for extracting spectra from JWST data cubes and interacting with the NASA PSG API, is available from https://github.com/apertureSynthesis/jwstComet. \texttt{jwstComet} makes use of the \texttt{astropy} \citep{astropy:2013, astropy:2018, astropy:2022}, \texttt{photutils} \citep{bradley2025}, and \texttt{astroquery} \citep{Ginsburg2019} libraries.

\item \textbf{Funding: }
Support for program \#5094 was provided by NASA through a grant from the Space Telescope Science Institute, which is operated by the Association of Universities for Research in Astronomy, Inc., under NASA contract NAS 5-03127. N.X.R., M.C., S.C. and S.M. were supported by the NASA Planetary Science Division Internal Scientist Funding Program through the Fundamental Laboratory Research work package (FLaRe). M.E.S. acknowledge support in part from UK Science and Technology Facilities Council (STFC) grant ST/X001253/1. D.F. conducted this research at the Jet Propulsion Laboratory, California Institute of Technology, under a contract with the National Aeronautics and Space Administration (80NM0018D0004).

\item \textbf{Author contributions:} 
N.X.R. performed the spectral extraction and modeling and led the manuscript writing. M.C. calibrated the JWST data. M.C, and G.V. contributed to the data analysis methodology and independently checked the molecular retrievals. M.M. performed astrometric measurements. D.F. calculated the 3I/ATLAS orbit and ephemeris. S.C. developed the formalism for estimating \ce{CH3D}/\ce{CH4} from $c-$\ce{C3HD}/$c-$\ce{C3H2} and helped interpret the isotopic ratios. N.B., D.B.-M., D.B., J.C., M.N.D., K.F., M.S.P.K., J.W.N., C.O. M.E.S., and C.A.S. helped with the project design, data acquisition, interpretation of results, and editing of the manuscript.

\item \textbf{Competing Interests:} The authors declare no competing interests.

\end{itemize}

\clearpage

\section*{Extended Data}

\begin{table}
\caption{\label{tab:dh} D/H ratios in solar system and extrasolar sources}
\centering
\begin{tabular}{cccc}
\hline\hline
Object & Molecule & D/H Ratio & Reference \\
\hline
ISM & \ce{H2} & $(2.0\pm0.1)\times10^{-5}$ & \cite{Prodanovic2010} \\
Protosolar & \ce{H2} & $(2.1\pm0.5)\times10^{-5}$ & \cite{Geiss1998} \\
Jupiter & \ce{CH4} & $(2.2\pm0.7)\times10^{-5}$ & \cite{Lellouch2001} \\
Saturn & \ce{CH4} & $(1.65\pm0.27)\times10^{-5}$ & \cite{Blake2021} \\
Earth & \ce{H2O} & $1.56\times10^{-4}$ & \cite{Marty2012} \\
Titan  & \ce{CH4} & $(1.59\pm0.33)\times10^{-4}$ & \cite{Nixon2012} \\
Uranus & \ce{CH4} & $(2.9\pm0.9)\times10^{-4}$ & \cite{Irwin2012} \\
Neptune & \ce{CH4} & $(3\pm1)\times10^{-4}$ & \cite{Irwin2014} \\
Makemake (ice) & \ce{CH4} & $(2.9\pm0.6)\times10^{-4}$ & \cite{Grundy2024} \\
Eris (ice) & \ce{CH4} & $(2.5\pm0.5)\times10^{-4}$ & \cite{Grundy2024} \\
Venus & \ce{CH4} & $1.25\times10^{-3}$ & \cite{Donahue1993} \\
67P/Churyumov-Gerasimenko & \ce{CH4} & $(2.41\pm0.29)\times10^{-4}$ & \cite{Mueller2022} \\
C/2007 W1 & \ce{CH4} & $<5\times10^{-3}$ $(3\sigma)$ & \cite{Bonev2009} \\
C/2001 Q4 & \ce{CH4} & $<1\times10^{-2}$ $(3\sigma)$ & \cite{Kawakita2005} \\
C/2007 N3 & \ce{CH4} & $<7.5\times10^{-3}$ $(3\sigma)$ & \cite{Gibb2012} \\
3I/ATLAS & \ce{CH4} & $(3.33\pm0.31)\times10^{-2}$ & This Work \\
IDP/UCAMM & Organics* & $9\times10^{-5}-8\times10^{-3}$ & \cite{Messenger2000} \\
 & & & \cite{Duprat2010} \\
Orgueil & Organic Radicals & $(1.5\pm0.5)\times10^{-2}$ & \cite{Gourier2008} \\
Disks & HCN* & $4\times10^{-3}-9\times10^{-2}$ & \cite{Oberg2012} \\
 & & & \cite{Huang2017} \\
IRAS 16293-2422B & \ce{H2CO} & $(3.25\pm1.00)\times10^{-2}$ & \cite{Persson2018} \\
                 & \ce{CH2DOH} & $(2.37\pm0.70)\times10^{-2}$ & \cite{Jorgensen2018} \\
                 & \ce{CH3OD} & $(3.2\pm0.9)\times10^{-3}$ & \cite{Ilyushin2024} \\
IRAS 2A & \ce{CH2DOH} & $1.93^{+0.43}_{-0.50}\times10^{-2}$ & \cite{Taquet2019} \\
        & \ce{CH3OD} & $7.10^{+1.20}_{-1.60}\times10^{-2}$ & \cite{Taquet2019} \\
IRAS 4A-NW & \ce{CH2DOH} & $1.23^{+0.40}_{-0.30}\times10^{-2}$ & \cite{Taquet2019} \\
           & \ce{CH3OD} & $3.10^{+1.00}_{-0.80}\times10^{-2}$ & \cite{Taquet2019} \\
HH212 & \ce{CH2DOH} & $(9.67\pm2.67)\times10^{-3}$ & \cite{Taquet2019} \\
      & \ce{CH3OD} & $(2.00\pm0.60)\times10^{-2}$ & \cite{Taquet2019} \\
V883 Ori & \ce{CH2DOH} & $(7.30\pm1.50)\times10^{-3}$ & \cite{Zeng2025} \\
         & \ce{CH3OD} & $(1.79\pm0.36)\times10^{-2}$ & \cite{Zeng2025} \\
HOPS 373-SW & \ce{CH2DOH} & $(4.4\pm1.1)\times10^{-2}$ & \cite{Lee2023} \\
L1527 & \ce{CH4}$^*$ & $(1.77\pm0.93)\times10^{-2}$ & \cite{Sakai2009} \\
Cha-MMS1 & \ce{CH4}$^*$ & $5.5\times10^{-2}$ & \cite{Lis2025} \\
Core 326/NGC1333 & \ce{CH4}$^*$ & $(4.69\pm0.72)\times10^{-2}$ & \cite{FerrerAsensio2026} \\
Perseus & \ce{CH4}$^*$ & $(0.25-4.5)\times10^{-2}$ & \cite{FerrerAsensio2026} \\
Cold Starless Cores & \ce{CH4}$^*$ & $(0.25-5.6)\times10^{-2}$ & \cite{Chantzos2018,FerrerAsensio2026} \\
HH211 & \ce{CH4}$^*$ & $(5.59\pm0.98)\times10^{-2}$ & \citep{Chantzos2018} \\
IRAS03282 & \ce{CH4}$^*$ & $(3.43\pm0.86)\times10^{-2}$ & \citep{Chantzos2018} \\
TMC-1 & \ce{CH4}$^*$ & $6.7\times10^{-2}$ & \cite{Gratier2016} \\
\hline
\end{tabular}
\parbox{12cm}{Note --- All D/H ratios are statistically corrected. The figures for organics in IDPs and UCAMMs, as well as for HCN in disks, are the ranges of values provided in the literature among multiple objects. Sources for which the D/H in $c-$\ce{C3H2} was used to estimate the D/H in \ce{CH4} are noted with an $*$ and the associated literature citations point to the publications reporting the D/H in $c-$\ce{C3H2}.}
\end{table}

\clearpage

\setcounter{figure}{0}
\setcounter{table}{0}

\renewcommand{\figurename}{Supplementary Data Figure} 
\renewcommand{\tablename}{Supplementary Data Table} 

\section*{Supplementary Figures}\label{figures}

\begin{figure*}[h]
\begin{center}
\includegraphics[width=\textwidth]{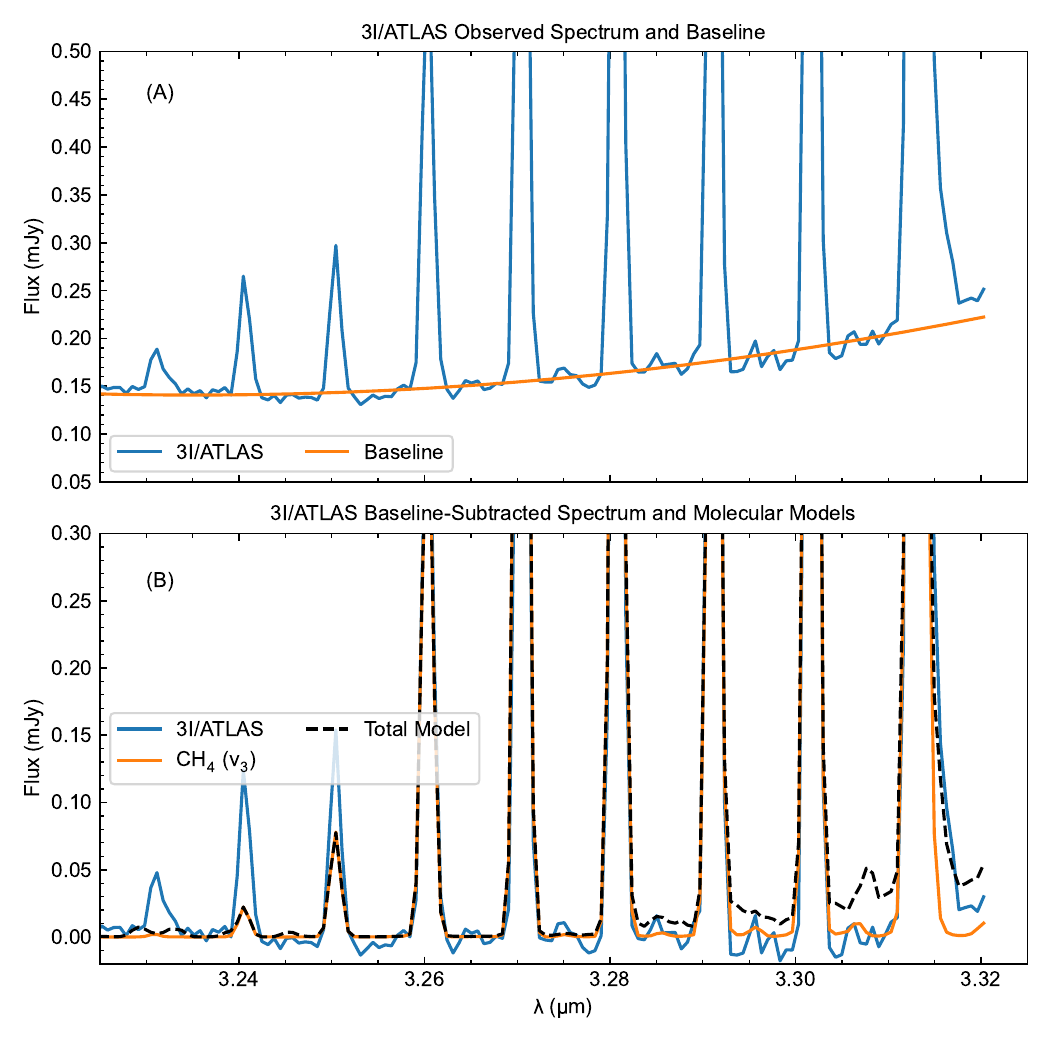}
\end{center}
\caption{\textbf{(A).} Observed 3I/ATLAS spectrum and second-order polynomial baseline, fit independently and before the fitting of PSG molecular emission models. \textbf{(B).} Best-fit PSG \ce{CH4} model fit to the baseline-subtracted spectra, demonstrating that a baseline fitted separately from the emission models does not provide improvement to the fits of the higher-$J$ \ce{CH4} lines (and simultaneously degrades the fit near the lower-$J$ lines), yet consistent $Q(\ce{CH4})$ and \trot{}(\ce{CH4}) are retrieved when considering only the lower-$J$ lines.}
\label{fig:baseline}
\end{figure*}

\begin{figure*}[h]
\begin{center}
\includegraphics[width=\textwidth]{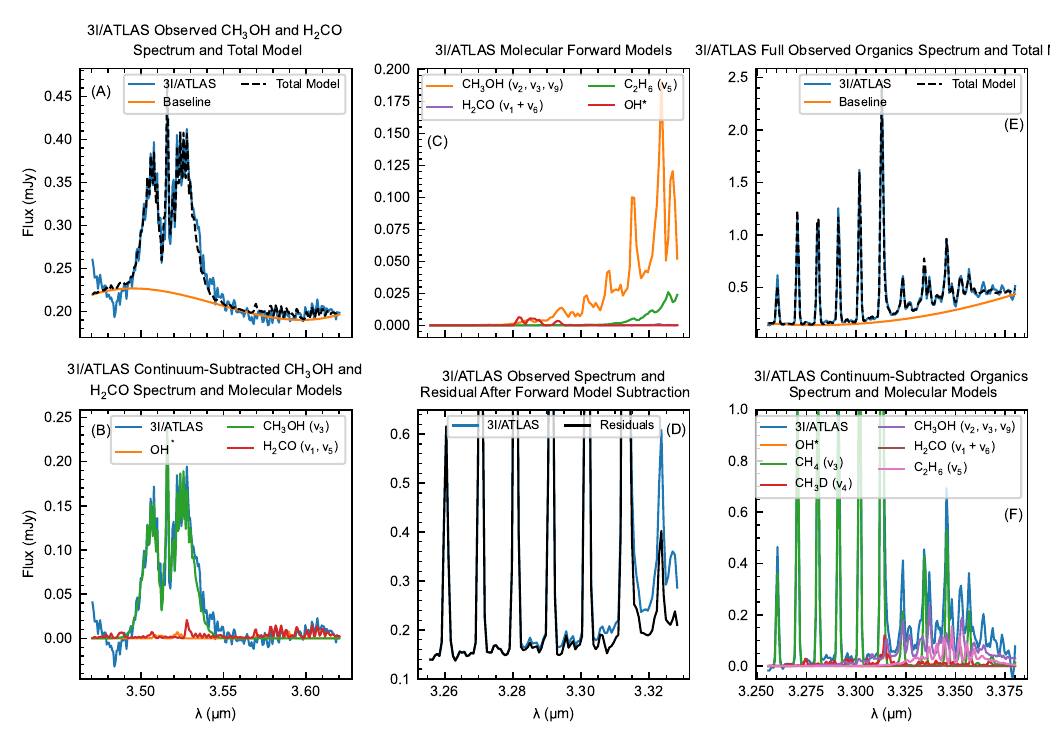}
\end{center}
\caption{\textbf{(A).} 3I/ATLAS \ce{CH3OH} and \ce{H2CO} spectrum on 2025 December 23 with best-fit model and spectral baseline overplotted. This was used to forward model both species when isolating \ce{C2H6}, \ce{CH4}, and \ce{CH3D}. \textbf{(B).} Baseline-subtracted 3I/ATLAS spectrum with best-fit individual molecular emission model(s) shown. \textbf{(C).} Forward modeled \ce{CH3OH}, \ce{C2H6}, \ce{H2CO}, and OH* spectra used to generate a \ce{CH4} $+$ \ce{CH3D} residual spectrum of 3I/ATLAS. \textbf{(D).} Comparison of observed 3I/ATLAS spectrum with the \ce{CH4} $+$ \ce{CH3D} residual spectrum generated after subtracting the forward models in the panel (C). \textbf{(E).} Observed 3I/ATLAS 3.3 $\mu$m organics spectrum with total emission model and baseline overplotted. \textbf{(F).} Baseline-subtracted 3I/ATLAS oganics spectrum with individual molecular emission models for all detected species shown.}
\label{fig:fits3}
\end{figure*}

\begin{figure*}[h]
\begin{center}
\includegraphics[width=\textwidth]{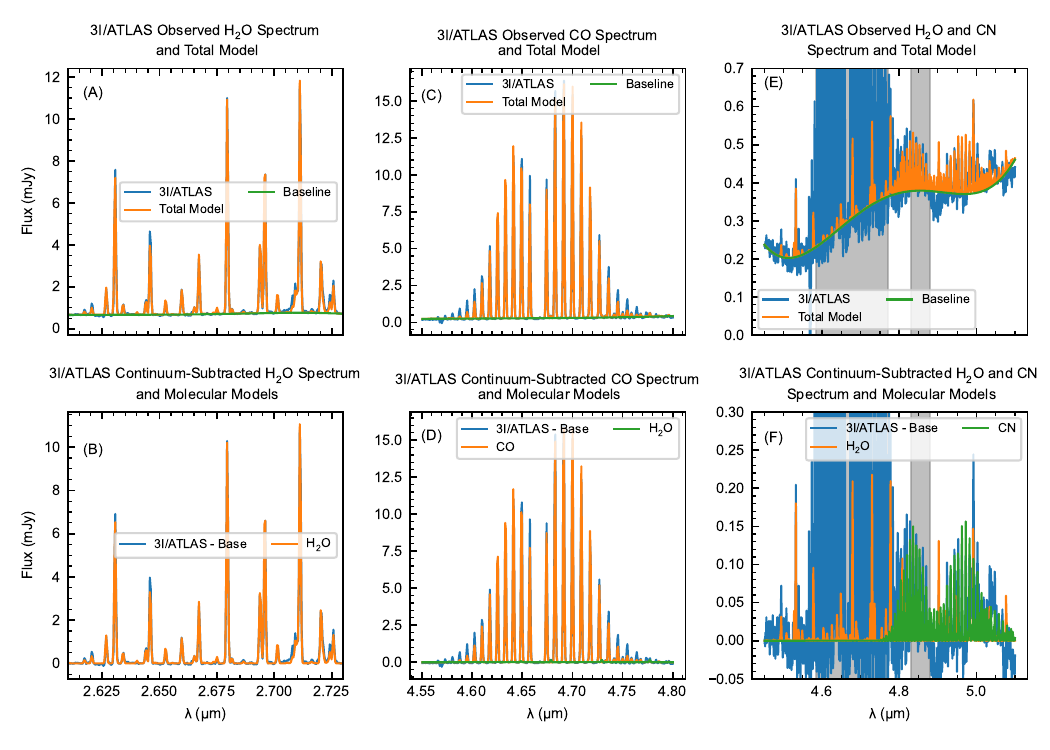}
\end{center}
\caption{\textbf{(A).} 3I/ATLAS 2.7 $\mu$m \ce{H2O} spectrum on 2025 December 22 with best-fit model and spectral baseline overplotted. \textbf{(B).} Baseline-subtracted 3I/ATLAS spectrum with best-fit individual molecular emission model(s) shown. \textbf{(C)-(D).} As in the left panels, but for CO on 2025 December 23. \textbf{(E)-(F).} As in the left panels, but for \ce{H2O} and CN on 2025 December 23. The gray shaded regions show the strong CO and OCS emission masked when retrieving $Q(\ce{H2O})$.}
\label{fig:fits2}
\end{figure*}

\section*{Supplementary Tables}\label{tables}

\begin{table}[h]
\caption{\label{tab:bands} Detected species and vibrational band identifications}
\centering
\begin{tabular}{ccc}
\hline\hline
Molecule & Vibration Band ID & $\lambda$ ($\mu$m) \\
\hline
CO & $\nu_1$ & 4.6 \\
\ce{H2O} & $\nu_1+\nu_3$ & 2.7 \\
       & $\nu1-\nu2$, $\nu_3-\nu_2$ & 4.5 \\
\ce{CH3OH} & $\nu_3$ & 3.5 \\
         & $\nu_2, \nu_9$ & 3.3 - 3.4 \\
\ce{H2CO} & $\nu_1, \nu_5$ & 3.5 \\
        & $\nu_1+\nu_6$ & 3.3 \\
\ce{C2H6} & $\nu_5$ & 3.35 \\
\ce{CH4} & $\nu_3$ & 3.25 \\
\ce{CH3D} & $\nu_4$ & 3.25 \\
\hline
\end{tabular}
\end{table}

\begin{table}[h]
\caption{\label{tab:baselines} Methane molecular production rates in 3I/ATLAS vs. spectral baseline}
\centering
\begin{tabular}{ccccc}
\hline\hline
Polynomial & $Q(\ce{CH4})$ & $Q(\ce{CH3D})$ & \trot{} & D/H Ratio \\
Order & $(10^{26}$ \ps{}) & $(10^{26}$ \ps{}) & (K) & (\%) \\
\hline
2 & $1.90\pm0.02$ & $0.25\pm0.02$ & $41.2\pm0.5$ & $3.33\pm0.31$ \\
3 & $1.90\pm0.02$ & $0.26\pm0.02$ & $41.2\pm0.5$ & $3.42\pm0.26$ \\
4 & $1.90\pm0.02$ & $0.26\pm0.02$ & $41.5\pm0.5$ & $3.45\pm0.26$ \\
\hline
\end{tabular}
\end{table}

\clearpage


\bibliography{3I}

\end{document}